\documentclass[conference]{IEEEtran}
\IEEEoverridecommandlockouts

\usepackage{cite}
\usepackage{amsmath,amssymb,amsfonts}
\usepackage{graphicx}
\usepackage[dvipsnames]{xcolor}
\usepackage{textcomp}
\usepackage[final]{microtype}
\usepackage[all]{nowidow}
\usepackage{balance}
\usepackage{booktabs}
\usepackage{url}
\usepackage[hidelinks]{hyperref}

\newcommand{\rev}[1]{#1}

\begin{document}

\bstctlcite{IEEEexample:BSTcontrol}

\title{TESSERA: A Workload-Driven Simulation and Design-Space Exploration
Framework for Heterogeneous NPUs}

\author{%
\IEEEauthorblockN{Arghadip Das$^{\dagger}$, Hoseok Kim$^{\dagger}$, Soomin Lee,
Arnab Raha, Deepak A Mathaikutty, and Vijay Raghunathan}
\thanks{$^{\dagger}$Equal contribution.}
\thanks{This work was supported in part by the Center for the Co-Design of
Cognitive Systems (CoCoSYS), one of the centers in the Joint University
Microelectronics Program (JUMP) 2.0, a Semiconductor Research Corporation
(SRC) program sponsored by the DARPA.}
\thanks{Arghadip Das (corresponding author, e-mail: das169@purdue.edu),
Hoseok Kim (e-mail: kim5396@purdue.edu), Soomin Lee (e-mail:
lee5782@purdue.edu), and Vijay Raghunathan (e-mail: vr@purdue.edu) are with
the Elmore Family School of Electrical and Computer Engineering, Purdue
University, West Lafayette, IN 47907 USA.}
\thanks{Arnab Raha (e-mail: arnab.raha@intel.com) and Deepak A Mathaikutty
(e-mail: deepak.a.mathaikutty@intel.com) are with the Advanced Architecture
Research Team, NPU IP, CGAI (CCG), Intel Corporation, Santa Clara,
CA 95054 USA.}
}

\maketitle

\begin{abstract}
AI model architectures are diversifying rapidly. Although dense
matrix multiplication still underlies today's CNNs and transformers,
emerging architectures (state-space models, long convolutions via
the fast Fourier transform (FFT), Kolmogorov--Arnold networks, and
spiking networks) are not multiply--accumulate (MAC) dominated; they
spend much of their computation on vector and non-MAC primitives
that homogeneous, MAC-centric neural processing units (NPUs) serve
poorly. This has motivated heterogeneous NPUs (HPUs) built from
non-identical tiles. Prior heterogeneous designs, however, vary only
one or two coarse knobs (typically MAC precision or array size) and
are evaluated on narrow workloads, and no existing framework
supports HPU design at a fine granularity,
where tiles differ across many architectural dimensions at once.
\looseness=-1

We present \textbf{TESSERA}, an analytical simulator and
design-space-exploration (DSE) framework for HPU
microarchitecture design. TESSERA searches the joint
space of tile-level heterogeneity: beyond array size and precision,
it varies the knobs along which tiles can differ, including
tile-type composition (large \emph{Big}, small \emph{Little}, and
non-MAC \emph{Special-Function} tiles), dataflow, sparsity mode, MAC engine
type, and special-function units for non-MAC operators (FFT,
spiking-integrate, polynomial). Unlike prior simulators, which model
a single homogeneous tile type, TESSERA models non-MAC
tiles with their own energy, area, and timing models and
maps operators across a mix of tiles with a heterogeneity-aware
compiler. A multi-seed pipeline that pairs a stratified sweep with
genetic-algorithm refinement returns Pareto-optimal designs, with
cost models calibrated to a 7\,nm node and cross-validated against
NVIDIA's Deep Learning Accelerator (NVDLA). Across a 20-workload
suite, the best general-purpose HPU found by
TESSERA ($\sim$200\,mm$^2$ Big$+$Little$+$Special-Function) achieves
\textbf{+46.91\%} mean iso-area energy savings over the best
iso-area homogeneous baseline. \looseness=-1

\end{abstract}


\begin{figure}[!t]
  \centering
  \includegraphics[width=\columnwidth]{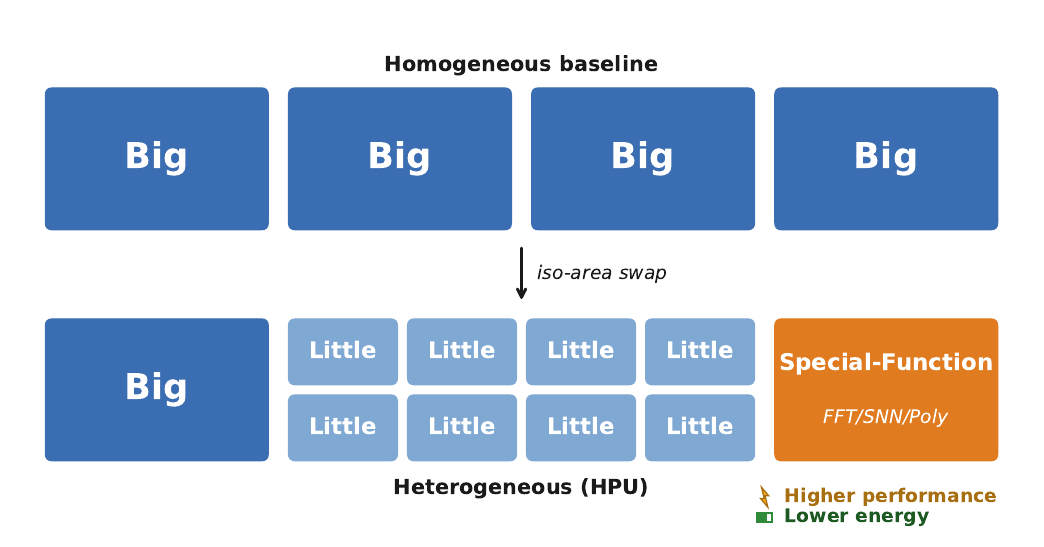}
  \caption{TESSERA's key idea: a homogeneous NPU replicates one
  MAC-centric tile (top); emerging non-MAC operators leave much of
  its silicon dark. At the \emph{same total area}, TESSERA's
  design-space exploration reallocates the chip into a heterogeneous
  mix of \emph{Big}, \emph{Little}, and \emph{Special-Function}
  (FFT/SNN/poly) tiles (bottom), yielding heterogeneous NPUs (HPUs)
  with +46.91\% mean iso-area energy savings over the best
  homogeneous baseline.}
  \label{fig:bg-iso-area-arch}
\end{figure}

\section{Introduction}\label{sec:introduction}

Neural-network workloads are diversifying rapidly, moving well beyond the dense general matrix multiplications (GEMMs) of convolutional neural networks (CNNs) and transformers~\cite{he2016deep,dosovitskiy2021image,touvron2023llama} to include selective state-space scans (Mamba~\cite{mamba}), long convolutions executed via the fast Fourier transform (FFT; Hyena~\cite{hyena}), Kolmogorov--Arnold networks built from polynomial basis functions (KAN~\cite{kan}), and spiking neural networks (SNNs)~\cite{snn}. A single multimodal inference pass now mixes many of these operator types at once. The hardware on which these workloads run, however, has not diversified at the same rate.

We refer to AI accelerators as neural processing units (NPUs); the term is generic, spanning the deployment spectrum from battery-powered edge and client devices to datacenter inference servers. Commercial NPUs from Intel~\cite{lnl}, Qualcomm~\cite{qualcomm_npu_whitepaper}, AMD~\cite{amd_xdna}, and MediaTek~\cite{mediatek_ai} replicate identical compute tiles built around large multiply--accumulate (MAC) arrays, vector digital signal processors (DSPs), and matched SRAMs, which are designs implicitly tuned for the dense GEMMs of CNNs and transformers. Emerging operators do not stress this substrate evenly. On commercial NPUs, MAC utilization varies sharply: dense CNNs saturate the array, while state-space-model (SSM) scans, FFT-based long convolutions, KAN polynomial bases, and graph-neural-network (GNN) gathers spend most of their time outside it~\cite{xamba,hkn,grannite}, so end-to-end inference latency degrades on hardware never built for them. The same mismatch leaves substantial \emph{dark silicon}. A quantized INT8 layer never touches the FP16 datapath, a depthwise convolution barely fills a $32{\times}32$ MAC array, and an FFT or polynomial operator's MAC-fabric lowering wastes most of the multiplier silicon.

These inefficiencies carry two costs, and both matter across the deployment spectrum. The primary cost is performance: emerging operators stall on a substrate built for dense GEMM, slowing inference on edge and datacenter NPUs alike. The secondary cost is energy: underused datapaths still draw power, which shortens battery life on edge devices and raises the operating cost and carbon footprint of datacenter inference.

An NPU is built from \emph{tiles}, and each tile contains one or more \emph{cores} (compute units such as a MAC array, a vector DSP, or a special-function unit) together with local memory and ports. Tile-level heterogeneity (i.e., composing a chip from non-identical tiles whose cores differ in type, precision, array size, dataflow, memory capacity, or specialized operator support) is one response to this mismatch, but no existing study or tool treats it as a primary design knob. Prior work shows that exposing architectural diversity improves latency, energy, or area efficiency for specific workload classes by varying MAC precision~\cite{spantidi_access}, array dimensions~\cite{maleki_hetero}, or coarse functional partitioning (a CNN core plus a recurrent-neural-network (RNN) core~\cite{dnpu_micro2018}, and plug-in coarse-grained reconfigurable arrays (CGRAs) for nonlinear large-language-model (LLM) operators~\cite{picachu_asplos2025}). However, each of these studies fixes a \emph{single coarse-grained knob} of heterogeneity and evaluates it on a \emph{narrow} workload spectrum. This coarse, hand-partitioned heterogeneity is now reaching commercial silicon: MediaTek's Dimensity 9500 (NPU 990) couples a high-performance compute engine with a low-power compute-in-memory (CIM) core in a dual-NPU design~\cite{dimensity9500_techinsights}. There is no tool today that lets an architect ask: \emph{which mix of tile types, MAC array dimensions, precisions, sparsity modes, dataflows, memory capacities, and special-function units is right for a given workload suite?} The simulators that enable design-space-exploration (DSE) (e.g., Timeloop~\cite{timeloop_ispass2019}, SCALE-Sim~\cite{scalesim}, MAESTRO~\cite{maestro}, and Voyager~\cite{voyager}) all assume a homogeneous tiling design, modeling a single uniform tile with no notion of mixed tile types or specialized non-MAC compute units. Consequently, no existing study or tool supports systematic exploration of multi-knob tile-level heterogeneity across diverse workloads.

We present \textbf{TESSERA}, an analytical simulator and DSE framework that designs heterogeneous NPUs (HPUs) by treating structural tile-level heterogeneity (Big, Little, and Special-Function (FFT/SNN/polynomial) tile types, each with a per-tile configurable precision set, composed per workload mix) as a joint design space. The two MAC-based types play complementary roles: a \emph{Big} tile is the general-purpose workhorse---a wide datapath that covers the broad operator mix at any precision from FP16 to INT4---whereas a \emph{Little} tile is a power-optimized, low-precision unit built for peak energy efficiency on the quantized layers that dominate modern inference. \emph{Special-Function} tiles add dedicated datapaths for the non-MAC operators (FFT, SNN, polynomial) that map poorly onto either. TESSERA jointly explores tile-type composition, MAC array geometry, precision (including INT4/INT8 quantization), sparsity mode, dataflow, memory-hierarchy capacity, and specialized functional units (FFT, SNN, polynomial). Given a target neural network suite, it produces per-operator and per-tile breakdowns of latency, energy, power, and utilization, together with tile-area estimates. These characterizations drive a multi-seed DSE pipeline (a stratified random sweep over the joint knob space followed by a per-area-budget \emph{genetic-algorithm (GA) refinement} seeded from the sweep bests) that returns Pareto-optimal architectures for the workload mix.

Our contributions are as follows:
\begin{itemize}
  \item \textbf{Heterogeneity-aware analytical simulator.} Unlike prior simulators, which model one homogeneous tile type, TESSERA models a chip as a mix of non-identical tiles, including non-MAC tiles for FFT, spiking-integrate, and polynomial operators, coordinated by a chip-level orchestrator that captures DRAM bandwidth sharing, cross-tile activation caching, network-on-chip (NoC) traffic, and per-tile power gating. (Section~\ref{sec:methodology-simulator})
  \item \textbf{Multi-seed DSE pipeline.} A stratified random sweep over the 12-knob design space (3 seeds, $\sim$2.94\,M samples), followed by per-area-budget GA refinement, searches the joint heterogeneous space and returns Pareto-optimal architectures. (Section~\ref{sec:methodology-dse})
  \item \textbf{20-workload characterization and three-group taxonomy.} The 20 workloads, spanning ten architectural families, fall into three groups by how much they benefit from heterogeneity: quantized LLMs/CNNs, FP16 transformer/SSM, and bandwidth-bound. (Section~\ref{sec:results-best})
  \item \textbf{Silicon-grounded calibration.} Per-module energy, area, and timing are calibrated against ASAP7 7\,nm synthesis~\cite{asap7} and cross-validated end-to-end against the open-source NVIDIA Deep Learning Accelerator (NVDLA)~\cite{nvdla} at two design points, backed by a system-level register-transfer-level (RTL) gating study. (Section~\ref{sec:methodology-calibration})
\end{itemize}

TESSERA's DSE returns HPUs that beat the best iso-area homogeneous designs across the suite. Peak savings reach $60.10 \pm 1.18\%$ on ResNet-50 (mean $\pm$ stdev, 3 seeds). The GA-refined Big$+$Little$+$Special-Function design at $\sim$200\,mm$^2$ reaches \textbf{+46.91\% mean iso-area energy savings} and \textbf{2.44 mean TOPS/W} (tera-operations per second per watt); a $\sim$100\,mm$^2$ variant is \textbf{1.5--2.4$\times$ faster} than NVDLA-large on INT8/SSM/ViT workloads. A complementary system-level RTL gating study delivers \textbf{28\% more MACs at 93.6\% lower power}.

\section{Background and Motivation} \label{sec:background}

\subsection{Generic NPU Tile}
\label{sec:bg-generic}
\begin{figure}[!t]
  \centering
  \includegraphics[width=\columnwidth]{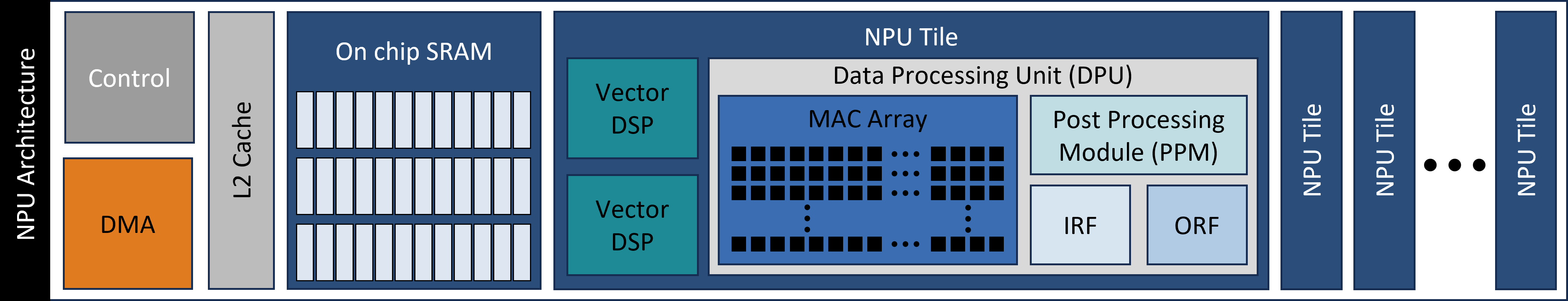}
  \caption{Generic NPU tile template: a MAC array and vector DSP fed
  from an SRAM scratchpad through input/output register files, with
  load/store ports to a shared DRAM channel.}
  \label{fig:bg-tile-template}
\end{figure}

Production NPUs are built by replicating an identical tile template
across the chip. Fig.~\ref{fig:bg-tile-template} shows the canonical
layout. Each tile pairs a MAC fabric (typically
a $16{\times}16$ to $64{\times}64$ systolic array supporting INT8 and
FP16) with a vector DSP that executes activations, normalizations,
elementwise operations, and softmax. A 256\,KB--2\,MB on-chip SRAM
scratchpad feeds the compute units through \rev{an input register
file (IRF) and an output register file (ORF)}, and dedicated
load/store ports stage tensors to and
from off-chip DRAM. Multiple tiles share a single DRAM channel and
communicate over a mesh, ring, bus, or NoC
interconnect. The resulting three-level energy hierarchy
($\sim$1--3\,pJ/byte at IRF/ORF, $\sim$5\,pJ/byte at SRAM,
40--200\,pJ/byte at DRAM~\cite{horowitz2014energy,muralimanohar2009cacti})
shapes every tiling, reuse, and dataflow decision an NPU compiler
makes.

Commercial NPUs from Intel~\cite{lnl}, Qualcomm~\cite{qualcomm_npu_whitepaper},
AMD~\cite{amd_xdna}, and MediaTek~\cite{mediatek_ai} all replicate this
template identically: every tile exposes the same precisions, the same
operator set, and the same memory hierarchy, all driven by a single
clock domain. The bet is that the tile is well-matched to a workload
mix dominated by dense convolution and matrix multiplication, and
that adding tiles is the right knob for scaling performance. The
remainder of this section examines whether that bet still holds.

\subsection{Operator Diversity in Modern Workloads}
\label{sec:bg-diverse-ops}

Neural-network workloads have moved past dense GEMM as the
single dominant kernel. A typical workload mix now spans dense CNNs
and vision transformers~\cite{he2016deep,dosovitskiy2021image} where
MAC fabrics still pay off, SSMs~\cite{mamba,s4} and
Hyena~\cite{hyena} long-convolution layers that lean on selective
scans and FFT, KANs~\cite{kan} that
evaluate polynomial bases per edge, SNNs that
integrate-and-fire, GNNs that scatter and gather,
hybrid attention/SSM LLMs (Nemotron-H~\cite{nemotronh},
Hymba~\cite{hymba_arxiv}) that interleave qualitatively different
layers within a single inference, and post-training-quantized LLMs
(LLaMA~\cite{touvron2023llama}, Mixtral~\cite{mixtral}, Nemotron~\cite{nemotronh} in INT4/INT8 forms) that ship with
$4\times$ higher arithmetic intensity than their FP16 counterparts.
Each family stresses a different subset of the tile.

\begin{figure}[!t]
  \centering
  \includegraphics[width=\columnwidth]{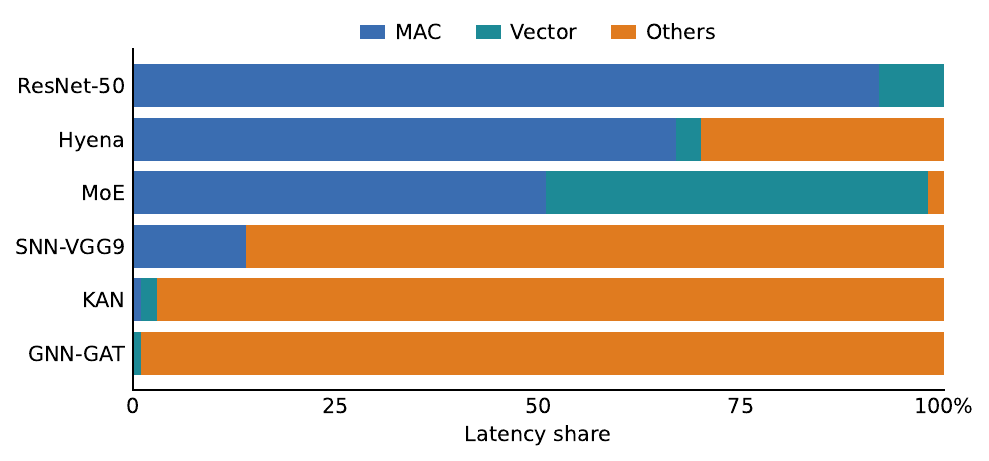}
  \caption{Per-operator inference latency breakdown on the Intel LNL
  NPU, grouped into MAC, Vector, and Other (non-MAC) operators. Only
  ResNet-50 is MAC-bound.}
  \label{fig:bg-latency-breakdown}
\end{figure}

\noindent\textbf{Quantitative evidence: end-to-end breakdown on real
silicon.} We measured per-operator latency for six representative
workloads on the Intel Lunar Lake (LNL) NPU~\cite{lnl} via OpenVINO's~\cite{openvino} per-op
profiling counters, grouping each operator into \emph{MAC}
(Conv/MatMul/FC), \emph{Vector} (elementwise, activation,
normalization, softmax; targeted at the DSP's single-instruction
multiple-data (SIMD) datapath), or \emph{Other}
(FFT, polynomial, SNN integrate, SSM scan, gather/scatter; no
dedicated hardware, lowered onto MAC or DSP).
Fig.~\ref{fig:bg-latency-breakdown} reports the breakdown.
ResNet-50 is the only MAC-bound workload. Hyena spends $\sim$30\% in
FFT lowered onto MAC; OLMoE's expert-routing softmax/gating
dominates the Vector share; SNN-VGG9 spends $\sim$47\% in leaky integrate-and-fire (LIF)
integration; KAN's wall time is entirely polynomial basis
evaluation; GNN-GAT is dominated by gather/scatter. Prior
measurement studies report MAC utilization below 10\% on commercial
NPUs for GNN~\cite{grannite}, Mamba/SSM~\cite{xamba}, and
KAN-style~\cite{hkn} networks. The dark-silicon problem is no
longer hypothetical; it is the common case for emerging model
families.

\subsection{Three Scopes of Heterogeneity}
\label{sec:bg-three-scopes}

Operator diversity manifests at three architectural scopes:
\emph{across models} (a deployed system-on-chip (SoC) serves FFT-heavy long
convolutions, LIF-bound SNNs, polynomial-bound KANs, INT4 LLMs,
and dense-GEMM CNNs that cannot share a single tile template),
\emph{within a model across layers} (artificial-neural-network/SNN
(ANN--SNN) networks alternate MAC and spiking-integrate layers;
Nemotron-H~\cite{nemotronh} and Hymba~\cite{hymba_arxiv} interleave
attention and SSM blocks), and \emph{within a layer across groups}
(group-level quantization, per-edge KAN bases of varying degree).
TESSERA targets the \emph{across-models} scope (one silicon
design per workload mix) and the \emph{across-layers} scope via
the mapper (Section~\ref{sec:methodology-compiler}); within-layer heterogeneity
is future work (Section~\ref{sec:future-work}). What is needed is a chip
composed of \emph{different} tile types tailored to the operator
mixture of the target workload suite: a \emph{heterogeneous}
architecture where tiles differ in MAC array geometry, supported
precisions, sparsity mode, dataflow, memory capacities, and the
presence or absence of dedicated functional units.

\subsection{The Need for a Heterogeneous Design Framework}
\label{sec:bg-need}

Prior heterogeneous-NPU studies have repeatedly demonstrated
the benefits of moving away from homogeneous tiling, but each fixes
a single coarse-grained knob and evaluates on a narrow workload
spectrum. DNPU~\cite{dnpu_micro2018} hard-wires a CNN core paired with
an RNN core; Spantidi et al.~\cite{spantidi_access} vary MAC precision
alone (the big.LITTLE NPU); Maleki et al.~\cite{maleki_hetero} vary
array dimensions alone; PICACHU~\cite{picachu_asplos2025} adds a
plug-in CGRA for nonlinear LLM operators. Production silicon is
moving the same way: MediaTek's Dimensity 9500 (TSMC N3P) ships a
dual-NPU SoC delivering 100\,TOPS~\cite{dimensity9500_techinsights},
making heterogeneous-NPU design a near-term commercial reality, not
just a research idea. Yet no prior simulator
supports cross-knob exploration: production DSE simulators
(Timeloop~\cite{timeloop_ispass2019}, SCALE-Sim~\cite{scalesim},
MAESTRO~\cite{maestro}, Voyager~\cite{voyager}) all model homogeneous
substrates with a single tile type and no dedicated models for
non-MAC accelerators (FFT, SNN, polynomial). An architect today
therefore has no principled way to ask: \emph{which mix of tile
types, precisions, MAC array geometries, sparsity modes, dataflows,
memory capacities, and special-function units is right for a given
workload suite?}

Closing this gap requires four co-designed capabilities that resist
adoption in isolation: (i) multi-tile heterogeneity (independent
tile types with the knobs above); (ii) dedicated energy/area/cycle
models for non-MAC accelerators; (iii) a heterogeneity-aware mapper
that filters operators by tile compatibility and schedules across
tiles with NoC-aware splitting; and (iv) a cross-knob search loop
over the joint design space. A hetero-aware mapper without non-MAC
models cannot evaluate Special-Function tiles; cross-knob search without
those prerequisites is uninformative. TESSERA is, to our
knowledge, the first framework to provide all four within a single
tool.

\rev{These four capabilities are tightly coupled, so retrofitting
them onto a homogeneous simulator would require abstractions that
such simulators lack. Supporting non-identical tile types changes
operator compatibility, cross-tile mapping and splitting, NoC
communication, shared DRAM-bandwidth arbitration, non-MAC cost
models, and the search loop itself. TESSERA therefore builds on
FlexNPU's per-tile analytical model~\cite{flexnpu} and adds the
heterogeneity-aware components of
Section~\ref{sec:methodology-compiler} through
Section~\ref{sec:methodology-dse}. TESSERA also stays at the
architecture-design stage and deliberately omits later-stage
features that existing tools already provide, namely exhaustive
kernel-level mapping, cycle-accurate traces, RTL generation,
quantization-accuracy evaluation, and
physical-design feedback. Those tools remain complementary
downstream.}

\subsection{Iso-Area Intuition}
\label{sec:bg-iso-area}

At a fixed chip area $N\times A$, replacing $N$ identical MAC tiles with
a heterogeneous mix of Big, Little, and optionally Special-Function tiles
extracts more performance and energy efficiency per mm$^2$ than
the homogeneous baseline. Fig.~\ref{fig:bg-iso-area-arch} sketches
the contrast. The top is the kind of homogeneous chip
today's commercial NPUs ship: $N$ identical FP16+INT8 MAC tiles
of footprint $A$ each; the bottom is a heterogeneous
reallocation: a smaller number of large Big tiles, several
smaller Little tiles, and optionally a Special-Function tile
whose primary compute is an FFT, SNN-integrate, or polynomial
unit rather than a MAC array, with total area held at $N\times A$. We
use ``Big'' and ``Little'' by analogy with CPU big.LITTLE~\cite{biglittle}: the
Big tile is large and typically carries the wider, higher-precision
datapath, the Little tile is small and typically restricted to a
narrower, lower-precision set---though the supported precisions are a
per-tile knob, not a fixed property of either type. This reallocation buys three things at the same $N\times A$:

\begin{itemize}
  \item \emph{Lower energy at the same area.} Lower-precision tiles
  consume less per MAC, and idle tiles can be
  power-gated to a small residual.
  \item \emph{Lower latency at the same area.} The freed area is
  reinvested into more parallelism on the tiles the workload
  actually exercises.
  \item \emph{Workloads that homogeneous cannot serve well.}
  Specialized functional units change the cost model
  asymptotically for FFT ($O(N\log N)$ butterflies vs.\ $O(N^2)$
  on a MAC-array lowering, a $\sim$100$\times$ blow-up at Hyena's
  typical $N{=}512$), and by large constant factors for LIF (a
  few gates per neuron vs.\ a multiplier-array lowering that
  contributes nothing to the result) and polynomial operators
  (KAN inference is reduced from a long multiply--accumulate
  chain hopping through SRAM at every step, to a $d$-cycle Horner-rule
  fused multiply-add pipeline with the accumulator pinned in a register), within
  the same area budget.
\end{itemize}

Which mix wins for which workload is precisely the question
TESSERA's DSE answers (Section~\ref{sec:results}).

\section{TESSERA Framework} \label{sec:methodology}

TESSERA takes a workload suite together with a candidate architecture and returns single-inference latency,
energy, area, power, and per-tile utilization. Driving the same
pipeline with a search loop produces Pareto-optimal heterogeneous
architectures over a $>10^{14}$-point design space.

\begin{figure*}[!t]
  \centering
  \includegraphics[width=\textwidth]{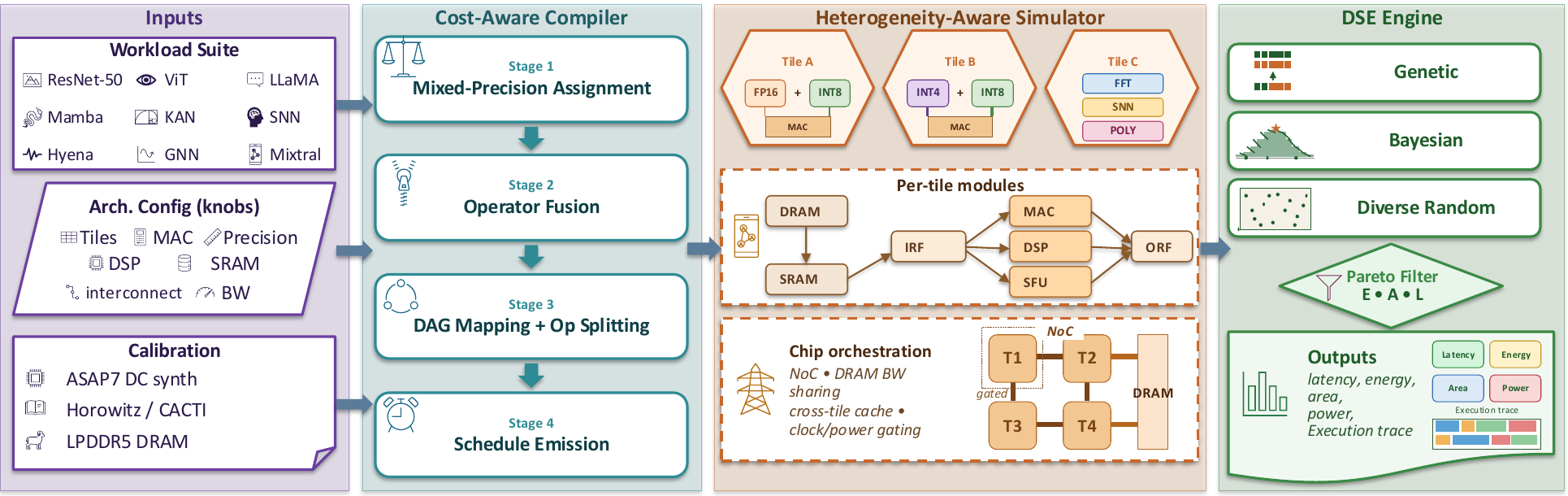}
  \caption{TESSERA framework overview: four cooperating
  layers---inputs, cost-aware compiler, heterogeneity-aware
  simulator, and silicon calibration---wrapped in a multi-seed DSE
  search loop.}
  \label{fig:methodology-overview}
\end{figure*}

Fig.~\ref{fig:methodology-overview} shows the four cooperating
layers: \emph{Inputs} (Section~\ref{sec:methodology-inputs}),
\emph{cost-aware compiler}
(Section~\ref{sec:methodology-compiler}; precision, fusion, mapping,
op-splitting, dataflow), \emph{heterogeneity-aware simulator}
(Section~\ref{sec:methodology-simulator}; per-tile module pipeline plus
chip-level orchestrator), \emph{calibration layer}
(Section~\ref{sec:methodology-calibration}; synthesized RTL + DRAM
literature), and the \emph{DSE engine}
(Section~\ref{sec:methodology-dse}; multi-seed stratified sweep + GA
refinement).

\subsection{Inputs}
\label{sec:methodology-inputs}

\noindent\textbf{Workload specification.}
A workload is a directed acyclic graph (DAG) of operators, each carrying a type (23-entry
vocabulary: 5 MAC-class, 15 DSP-class, 3 special (SNN-integrate,
FFT, polynomial)), shape, precision (INT4/INT8/FP16/BF16/FP32), and
per-operand sparsity rates. Workloads are imported from ONNX~\cite{onnx} or
extracted from PyTorch~\cite{paszke2019pytorch}.

\noindent\textbf{Architecture configuration.}
An architecture lists one or more \emph{tile templates}, per-template
instance counts, an interconnect topology (mesh/bus/ring/NoC), DRAM
size, bandwidth, and latency. Each tile template exposes a small
set of architectural knobs: enabled modules (MAC array, DSP, special
unit, load/store ports), MAC engine type and dimensions, supported
precisions, sparsity mode (none, activation-/weight-sided,
two-sided, or structured $N{:}M$~\cite{flexcim_tvlsi2026}),
SRAM capacity and banking, IRF/ORF granularity, DSP unit count and
SIMD width, and double-buffering.
Each tile type is clocked in its own fixed domain. The same schema
describes a homogeneous chip (one template), a mixed-precision chip
with two tile types, or a three-tile-type chip (e.g., an FP16+INT8 Big,
an INT4+INT8 Little, and a Special-Function tile). A representative homogeneous
baseline mirroring an Intel LNL-class NPU~\cite{lnl} is expressed as
a set of identical FP16+INT8 MAC tiles with matched SRAM and DSP
units, sharing a mesh interconnect and a single DRAM channel.

\subsection{Cost-Aware Compiler}
\label{sec:methodology-compiler}

The compiler converts a (workload, architecture) pair into an
execution plan in four ordered passes. Each pass tags operators
for the simulator and DSE; no machine code is emitted.

\noindent\textbf{(1) Mixed-precision assignment.} Default policy
(Conv/MatMul/Pool$\to$INT8;
LayerNorm/RMSNorm/Softmax/SNN/FFT/polynomial/SSM scan$\to$FP16) with
a name-based override forcing FP16 on accuracy-sensitive layers
(attention QKV/output projection, LM head, classifier,
embedding)~\cite{llm_npu}.
An \emph{aggressive} mode demotes all convolutions to INT4.

\noindent\textbf{(2) Operator fusion.} A greedy left-to-right scan
matches three-op (Conv+BN+Act, Conv+Add+Act) and two-op patterns;
matched groups fold post-processing into the tile's post-processing
module (PPM), skipping the SRAM round-trip for intermediate tensors.

\noindent\textbf{(3) DAG-aware mapping with op-splitting.}
Operators are visited in topological order. For each operator $o$,
the mapper filters tiles by op-type and precision compatibility,
then for each compatible tile $T$ computes an earliest start time
\begin{equation}
\begin{split}
t_{\mathrm{start}}(o,T) = \max\Big(\,&\mathrm{tile\_finish}[T],\\
&\max_{(f_j,T_j)\in\mathrm{preds}(o)}\!\big(f_j+\mathbf{1}[T_j{\neq}T]\,\Delta_{\mathrm{NoC}}\big)\!\Big),
\end{split}
\label{eq:earliest-start}
\end{equation}
\rev{where $\mathrm{tile\_finish}[T]$ is the cycle at which $T$
finishes the work already scheduled on it, and the second term is
the latest time at which a predecessor's output is available on $T$. Predecessor $j$ finishes at cycle $f_j$ on its assigned tile
$T_j$, and the indicator $\mathbf{1}[T_j{\neq}T]$ adds the NoC
delay $\Delta_{\mathrm{NoC}}$ only when $j$ ran on a different
tile. The mapper then computes}
a roofline cycle estimate (the larger of the compute-bound
and bandwidth-bound cycle counts)
\begin{equation}
\hat C_{\mathrm{op}}(T) = \max\!\Big(
  \big\lceil \tfrac{\mathrm{MACs}}{R_T C_T \eta_T}\big\rceil,\;
  \big\lceil \tfrac{\mathrm{bytes}}{B_{\mathrm{DRAM}}(T)}\big\rceil
\Big),
\label{eq:roofline}
\end{equation}
where \rev{$\mathrm{MACs}$ and $\mathrm{bytes}$ are the operator's
MAC count and DRAM traffic, $B_{\mathrm{DRAM}}(T)$ is the DRAM
bandwidth available to tile $T$ in bytes per cycle,}
$R_T{\times}C_T$ is the MAC array shape\rev{,} and $\eta_T$ is a
per-MAC throughput multiplier ($>{1}$ when sparsity skipping
applies). The mapper places $o$ on the tile minimizing
$t_{\mathrm{start}}+\hat C_{\mathrm{op}}$ (completion time, not just
cycle estimate). For MAC-class ops with multiple compatible tiles,
it evaluates an even split along the output-channel ($OC$), batch
($B$), or input-channel ($IC$) dimension with explicit reduce/concat
cost
\begin{equation}
C_{\mathrm{reduce}} =
  \max_i\big(\lceil B^{\mathrm{out}}_i / B_{\mathrm{NoC}} \rceil + \Delta_{\mathrm{NoC}}\big),
\label{eq:reduce}
\end{equation}
\rev{where $B^{\mathrm{out}}_i$ is the byte count of partition
$i$'s output returned over the NoC (a partial-sum reduction for
$IC$ splits, a concatenation for $OC$/$B$ splits) and
$B_{\mathrm{NoC}}$ is the NoC bandwidth in bytes per cycle,}
accepting the split only if its finish time beats single-tile
placement. \rev{When the split dimension does not divide evenly
across the participating tiles, each tile receives a whole number
of elements along that dimension and the final partition absorbs
the remainder, so the partitions are unequal in size. Each
partition is then costed on its assigned tile with
Eqs.~\eqref{eq:earliest-start}--\eqref{eq:reduce} before the split
is accepted.} Dataflow (weight-stationary (WS), output-stationary
(OS), row-stationary (RS), or AUTO) is picked per operator\rev{,
because the best dataflow for a matmul depends on its shape and no
single fixed dataflow such as WS is best across all shapes. In an
$M{\times}K$ by $K{\times}N$ matmul the operands are sized
$M{\cdot}K$ and $K{\cdot}N$ and the output $M{\cdot}N$. OS holds
the $M{\cdot}N$ partial sums in the array and accumulates them in
place, avoiding repeated partial-sum traffic to SRAM, so OS
pays off only when the output is far larger than the operands. 
WS is the default, holding the $K{\cdot}N$ weights and reusing them
across the $M$ activation rows.} AUTO \rev{therefore}
chooses OS when $M{\cdot}N$ exceeds both $K{\cdot}N$ and
$M{\cdot}K$ by $4{\times}$, WS otherwise. \rev{The $4{\times}$
threshold is a heuristic rather than an analytically derived
value, set large enough that AUTO stays on WS unless the output
clearly dominates.}

\noindent\emph{What changes under a heterogeneous architecture.}
The compatibility filter routes each op to the smallest compatible
tile (FP16 MATMUL$\to$Big, INT8 Conv$\to$any, FFT$\to$Special-Function);
splittable MAC ops are partitioned across Big$+$Little so the bulk
runs on the small tiles. FP16-only ops on chips with one
FP16-capable tile serialize.

\noindent\textbf{(4) Schedule emission.} A scheduler converts the
per-op mapping into an execution schedule (latency mode parallelizes
distinct-tile assignments; throughput mode pipelines multi-batch).

\subsection{Heterogeneity-Aware Simulator}
\label{sec:methodology-simulator}

The simulator consumes the compiled execution plan, the architecture
configuration, and the calibration tables and returns latency,
energy, area, and per-tile utilization. It is built on the
FlexNPU~\cite{flexnpu} dataflow-aware analytical model, extended with
non-MAC accelerator modules and a chip-level orchestrator that handles
heterogeneous tile mixes. The tile-template plus NoC abstraction is
general enough to model not just single-NPU multi-tile designs but
also multi-NPU SoCs such as MediaTek's dual-NPU
Dimensity 9500~\cite{dimensity9500_techinsights}, where each NPU
instance is expressed as a tile cluster sharing an inter-NPU NoC
channel.

\subsubsection{Per-Tile Module Pipeline}
\label{sec:methodology-tile-pipeline}

A tile is modeled as seven hardware modules; three are compute cores (the MAC array, the DSP, and the special-function unit) and the rest handle memory and data staging. Given a compiled
operator, the tile simulator routes the operator through one of
three execution paths (MAC, DSP, or Special-Function) and accumulates
cycles and energy at each module.

\begin{itemize}
  \item \emph{Compute module (MAC array).} An SRAM-budget tiling
  pass decomposes the operator along $(M,K,N)$ into tiles that fit
  the working set (weights and double-buffered activations within
  the per-tile SRAM). Per-tile cycles follow an engine-specific model;
  for systolic arrays,
  \begin{equation}
    C_{\mathrm{sys}} = \sum_{n,k}\!\Big[D +
      \sum_m (m_{\mathrm{eff}}+k_{\mathrm{eff}}+D-2)\Big],
    \label{eq:systolic-cycles}
  \end{equation}
  \rev{where $m$, $k$, $n$ index the $(M,K,N)$ tiles produced by
  the tiling pass, $m_{\mathrm{eff}}$ and $k_{\mathrm{eff}}$ are
  the effective $M$ and $K$ extents of the current tile (smaller at
  boundary tiles), and $D$ is the systolic pipeline depth, so each
  tile pays $m_{\mathrm{eff}}{+}k_{\mathrm{eff}}{+}D{-}2$ cycles of
  array fill and drain.} Compute energy is sparsity-aware MAC count
  times per-MAC energy from the calibration table; six sparsity modes
  are supported.
  \item \emph{DRAM module.} Burst-aligned reads/writes at per-tile
  bandwidth, $E_{\mathrm{DRAM}}=(\mathrm{bytes_{rd}}+\mathrm{bytes_{wr}})\,E_{\mathrm{DRAM/B}}$.
  \item \emph{SRAM module.} Tiling-aware reuse from the chosen
  dataflow (WS/OS/RS), with grid heuristics as fallback.
  \item \emph{IRF / ORF.} IRF pays write energy padded to write
  granularity, reads reduced by activation sparsity. ORF is
  $K$-tile-aware (first $K$-tile write-only, later ones read-modify-write).
  \item \emph{DSP module.} A SIMD instruction set of 14 vector operations (\texttt{vadd},
  \texttt{vmul}, \texttt{vexp}, \texttt{vreduce}, \texttt{vlut}, \ldots)
  covers every elementwise/normalization/activation/softmax operator;
  each high-level op decomposes into a vector sequence. SSM scan
  carries a sequence-length sequential multiplier.
  \item \emph{Special-Function module (special-function unit, SFU).} Dedicated formulas for radix-2
  FFT ($N\log_2 N$ cycles \rev{for an $N$-point transform}), LIF
  ($\lceil N/N_{\mathrm{par}}\rceil T$ cycles \rev{for $N$ neurons
  integrated over $T$ timesteps on $N_{\mathrm{par}}$ parallel
  lanes}), and Horner polynomial ($Nd$ cycles \rev{for $N$
  evaluations of a degree-$d$ polynomial}).
\end{itemize}

\subsubsection{Total-Cycle Model}
\label{sec:methodology-dataflow-walkthrough}

The seven-module pipeline interleaves DRAM staging, SRAM
tiling-aware reuse, IRF/ORF activation/partial-sum feed, and
PPM post-processing on top of each tile's compute path. With
double-buffering (DB), total cycles overlap compute, memory, and
DRAM:
\begin{equation}
C_{\mathrm{tot}} =
\begin{cases}
  \max(C_{\mathrm{cmp}},C_{\mathrm{mem}},C_{\mathrm{DRAM}})
    + C_{\mathrm{LP}}+C_{\mathrm{SP}} & \text{DB}\\
  C_{\mathrm{cmp}}+C_{\mathrm{mem}}+C_{\mathrm{DRAM}}
    +C_{\mathrm{LP}}+C_{\mathrm{SP}} & \text{no DB}
\end{cases}
\label{eq:total-cycles}
\end{equation}
where \rev{$C_{\mathrm{cmp}}$ is the cycle count of the operator's
compute path (MAC, DSP, or SFU), $C_{\mathrm{mem}}$ the on-chip
SRAM and IRF/ORF staging cycles, $C_{\mathrm{DRAM}}$ the off-chip
transfer cycles, and} $C_{\mathrm{LP}}$, $C_{\mathrm{SP}}$ are load/store-port direct-memory-access (DMA)
cycles.

\subsubsection{Latency, Energy, and Area Models}
\label{sec:methodology-models}

\noindent\textbf{Latency.} Per-tile latency $L=C_{\mathrm{tot}}/f$;
end-to-end latency is the makespan of the chip schedule including
NoC delays and split-op reductions.

\noindent\textbf{Energy.} Total tile energy sums contributions from
every enabled module:
\begin{equation}
\begin{split}
E_{\mathrm{tile}} = {}&E_{\mathrm{cmp}}+E_{\mathrm{DRAM}}+E_{\mathrm{SRAM}}+E_{\mathrm{IRF}}\\
&{}+E_{\mathrm{ORF}}+E_{\mathrm{DSP}}+E_{\mathrm{spec}}-E_{\mathrm{fuse}},
\end{split}
\label{eq:tile-energy}
\end{equation}
where \rev{each right-hand term is the calibrated energy of the
corresponding module and} $E_{\mathrm{fuse}}=N_{\mathrm{fused}}\cdot2\,|\mathrm{out}|
\,E_{\mathrm{SRAM/B}}$ accounts for SRAM round-trips avoided by
fused intermediates\rev{. Each of the $N_{\mathrm{fused}}$ fused
pairs skips one write and one read of an $|\mathrm{out}|$-byte
intermediate tensor at $E_{\mathrm{SRAM/B}}$ per byte}. Chip energy adds NoC transfer and static
leakage from power-gated tiles.

\noindent\textbf{Area.} Tile area is the analytical sum
\begin{equation}
\begin{split}
A_{\mathrm{tile}} = {}&
N_{\mathrm{MAC}}\cdot \max_p A_{\mathrm{MAC}}(p)\\
&+ A_{\mathrm{SRAM}}+A_{\mathrm{DSP}}+A_{\mathrm{spec}}+A_{\mathrm{ports}},
\end{split}
\label{eq:tile-area}
\end{equation}
with \rev{$N_{\mathrm{MAC}}=R_T C_T$ the tile's MAC count and}
per-MAC area \rev{$A_{\mathrm{MAC}}(p)$} taken over the largest
supported precision \rev{$p$}
(multi-precision MACs include the wide datapath); IRF/ORF area
folds into $A_{\mathrm{ports}}$. Chip area is
$A_{\mathrm{chip}}=\sum_t A_{\mathrm{tile}}(t)+A_{\mathrm{NoC}}$,
omitting floorplan dead space (flagged in Section~\ref{sec:future-work};
RTL gating study in Section~\ref{sec:results-validation} bounds it).

\subsubsection{Heterogeneous Mechanisms}
\label{sec:methodology-hetero-mech}

Two mechanisms make the analytical model heterogeneity-aware.

\noindent\emph{Dynamic DRAM bandwidth sharing.} Only tiles whose
previous operator has not yet finished are counted active; per-tile
bandwidth is $\mathrm{BW}_{\mathrm{total}}/N_{\mathrm{active}}$, so
a subset of busy tiles gets full bandwidth rather than a static
share.

\noindent\emph{Cross-tile activation caching and synchronization.}
Each tile's SRAM splits between a working set and a FIFO-evicted
activation cache. Three cases: local cache hit (same-tile
producer/consumer $\Rightarrow$ no DRAM read), cross-tile DMA
(NoC transfer at $\lceil B/B_{\mathrm{NoC}}\rceil$ ${+}\,h\,C_{\mathrm{base}}$
cycles \rev{for $B$ bytes moved at NoC bandwidth $B_{\mathrm{NoC}}$
bytes per cycle over $h$ hops with per-hop base latency
$C_{\mathrm{base}}$}), or cache miss (full DRAM load). A pre-built consumer map
enforces dependency synchronization across heterogeneous tiles.

\noindent\emph{Clock and power gating.} Idle modules within an active
tile are clock-gated (no dynamic energy). Tiles with no scheduled
work are power-gated at 5\% residual leakage, a figure consistent
with header-switch power gating at modern technology nodes.

\subsubsection{Heterogeneous Tile Types}
\label{sec:methodology-tile-types}

The tile-template knobs let a single configuration schema represent
three qualitatively different tile types that recur in our DSE
outputs. The supported-precision set is itself a per-tile knob: any
tile may declare an arbitrary subset of the available datatypes, so
the precision labels below are illustrative of representative DSE
points, not hard-wired tile classes.
\begin{itemize}
  \item \emph{Big tile.} A large systolic array with
  ample SRAM, two-sided sparsity, and dual DSP. Hosts the
  highest-precision operators in a configuration---e.g., FP16
  attention, normalization, and high-accuracy convolution---and is
  typically provisioned with a wide precision set (e.g., FP16+INT8).
  \item \emph{Little tile.} A small array with modest
  SRAM, a single DSP, and optional sparsity. Hosts the
  lower-precision, INT-quantizable convolution and matmul at
  substantially lower energy than the Big tile, and is typically
  provisioned with a narrower, lower-precision set (e.g., INT4+INT8).
  \item \emph{Special-Function tile.} No MAC array. Includes one or more SFUs
  (FFT, SNN-integrate, polynomial), modest SRAM, and a single DSP for
  pre/post-processing. Hosts the non-MAC operator classes for which
  MAC lowering is asymptotically wasteful.
\end{itemize}
The same chip simulator orchestrates any cross product of these tile
types; what differs across architectures is the per-type count and
per-type sizing.

\subsubsection{Outputs}
\label{sec:methodology-outputs}

For each (workload, architecture) pair the simulator emits
end-to-end latency, total/average power, and area; per-tile active
cycles, achieved/peak TOPS, TOPS/W, TOPS/mm$^2$, arithmetic
intensity, and roofline class; per-module energy breakdowns
(compute, DRAM, SRAM, IRF, ORF, DSP, special, NoC, leakage); and an
execution trace in the Perfetto profiling format for visual inspection of tile
utilization and cross-tile movement.

\subsection{Calibration}
\label{sec:methodology-calibration}

Energy, area, and timing parameters come from Synopsys Design
Compiler (DC) synthesis of every RTL module at the
ASAP7 7\,nm process design kit~\cite{asap7}, 2\,GHz target. SRAM area is taken
from CACTI~7.0~\cite{muralimanohar2009cacti}; SRAM leakage is
taken from DC.

\noindent\textbf{DRAM calibration.} DRAM energy and bandwidth come
from DRAM-process literature rather than logic synthesis. The
ASAP7 calibration pairs with LPDDR5-6400 (40\,pJ/byte, 51.2\,GB/s
rounded to 64\,GB/s on the DSE grid, 100-cycle access latency). The
bandwidth is a DSE knob and is swept over
\{16, 32, 64, 128, 256, 512\}\,GB/s.

\noindent\textbf{System-level RTL validation.} A dedicated study
packages a homogeneous two-tile system and an iso-area heterogeneous
(one FP16+INT8 + one INT4+INT8) system into 109 SystemVerilog files
($\sim$19.3k lines of code) synthesized at ASAP7. The homogeneous design
clock-gates the FP16 datapath when running INT8; the heterogeneous
design power-gates the INT4+INT8 tile when idle. Results in
Section~\ref{sec:results-validation}.

\noindent\textbf{External cross-validation.} The same calibration
flow matches TESSERA against NVIDIA NVDLA~\cite{nvdla} at two design points
spanning $32{\times}$ in MAC density: \texttt{nv\_small} (8$\times$8
INT8 systolic, 64\,KB convolution buffer (CBUF)) and \texttt{nv\_full} (32$\times$64
INT8+FP16, 512\,KB CBUF), using the published NVDLA Primer's area
and energy values as the external reference. NVDLA is the natural
target here, and effectively the only one: it is the sole openly
available production NPU that ships synthesizable RTL together with a
published per-module area and energy breakdown, which is why
analytical DSE simulators are likewise each validated against a
single such reference (e.g., Timeloop~\cite{timeloop_ispass2019}
against the Eyeriss accelerator). NVDLA is an edge-class design, so this check exercises the framework at edge area and power scales; the same cost models and search apply to datacenter-scale NPUs, where only the area, power, and bandwidth budgets supplied to the DSE differ.

\subsection{Design Space Exploration}
\label{sec:methodology-dse}

The DSE engine drives the compiler+simulator over the joint
heterogeneous knob space and returns a Pareto front in (energy,
area, latency) together with the best design under a user-supplied
scalar objective.

\noindent\textbf{Knobs.}
The DSE varies 12 knobs spanning tile-type composition, MAC-array
geometry, precision, on-chip memory, DRAM bandwidth, sparsity,
dataflow, interconnect, and pipelining. The full knob space exceeds
$10^{14}$ configurations; the per-knob value grid is enumerated in
Section~\ref{sec:expt-dse}.

\noindent\textbf{Search pipeline.} The DSE runs in two stages,
across multiple random seeds for confidence intervals.
\emph{Sweep:} stratified random sampling over the knob space
(strata = area budget $\times$ architecture family); it is the
source of the per-workload best-of-DSE numbers in
Section~\ref{sec:results-dse}. \emph{GA refinement:} a per-area-budget
genetic algorithm seeded from the best sweep individuals,
returning the best general-purpose heterogeneous architecture per
area budget (Section~\ref{sec:results-ga}). Alongside these two stages, the DSE engine
also provides a Bayesian optimization (BO) search backend (a surrogate
model fit to evaluated configurations, with new candidates proposed
by an acquisition function) as a sample-efficient alternative to the
GA refinement when the simulation budget is constrained.

\noindent\textbf{Objective.} The GA-refinement fitness is the
workload-equal-weighted mean iso-area energy savings of the
candidate design over the best homogeneous design at the same area
(from the sweep), normalized by the best peak TOPS/W observed:
\begin{equation}
\mathrm{fitness}(d) = \overline{\Delta E}_{\mathrm{iso\text{-}area}}(d)
                       \;+\; \alpha \cdot \tfrac{\mathrm{TOPS/W}(d)}{\max_{d'} \mathrm{TOPS/W}(d')},
\label{eq:dse-objective}
\end{equation}
where \rev{$\overline{\Delta E}_{\mathrm{iso\text{-}area}}(d)$ is
the mean iso-area energy savings of candidate design $d$ and}
$\alpha$ is a small positive tie-breaker. We report both
\emph{peak} TOPS/W (best-single workload) and \emph{mean} TOPS/W
(workload-equal-weighted); for general-purpose chip-level claims
the mean is the right metric.

\section{Experimental Methodology} \label{sec:expt}

\subsection{Workload Suite}
\label{sec:expt-workloads}

We evaluate TESSERA on a 20-workload suite: 14 base models
spanning ten architectural families, plus six post-training-quantized
INT4/INT8 variants of the widely deployed transformer LLMs
(Table~\ref{tab:workload-suite}). The suite was constructed to
(i) exercise all 23 operator types in TESSERA's vocabulary,
(ii) stress every execution path on the tile (MAC, DSP, Special-Function),
(iii) span five orders of magnitude in arithmetic intensity, and
(iv) cover the INT4/INT8 post-training quantization that
production LLMs increasingly ship in.

\begin{table}[!t]
\centering
\caption{Workload suite. 14 base models in ten architectural
families; transformer LLMs include INT4/INT8 quantized variants.}
\label{tab:workload-suite}
\setlength{\tabcolsep}{4pt}
\footnotesize
\begin{tabular}{@{}lll@{}}
\toprule
Family                       & Model              & Precision \\
\midrule
CNN                          & ResNet-50~\cite{he2016deep} & INT8 \\
Vision Transformer           & ViT-B/16~\cite{dosovitskiy2021image} & FP16, INT8 \\
Dense LLM                    & LLaMA-7B~\cite{touvron2023llama} & FP16, INT8, INT4 \\
Dense LLM                    & Spec.\ decode~\cite{specdecode} & FP16 \\
Mixture-of-experts LLM       & Mixtral~\cite{mixtral} & FP16, INT4 \\
Hybrid attention/SSM LLM     & Nemotron-H~\cite{nemotronh} & FP16, INT8, INT4 \\
SSM                          & Mamba-370M~\cite{mamba} & FP16 \\
SSM                          & Hyena-1.3B~\cite{hyena} & FP16 \\
KAN                          & KAN~\cite{kan} & FP16 \\
SNN                          & SNN-VGG9~\cite{snn} & FP16 \\
Multimodal                   & LAVISH~\cite{lavish} & FP16 \\
Multimodal                   & LLaVA~\cite{llava} & FP16 \\
Multimodal                   & RT-2~\cite{rt2} & FP16 \\
GNN                          & GNN-GAT~\cite{gat} & FP16 \\
\bottomrule
\end{tabular}
\end{table}

\subsection{Metrics}
\label{sec:expt-metrics}

We report \rev{end-to-end single-batch inference latency}, energy,
and area as the primary \rev{metrics} (lower is better)\rev{,
consistent with the Pareto objectives in
Section~\ref{sec:methodology-dse}}. Energy \rev{is the}
single-inference energy with per-module breakdown (compute, DRAM,
SRAM, IRF, ORF, DSP, special, NoC) split into dynamic vs.\ static.
Area \rev{is the} total chip area (mm$^2$) with per-tile breakdown into MAC
array, SRAM, DSP, special unit, and ports. \rev{Achieved TOPS,
per-tile processing-element utilization, and the derived}
\emph{peak} TOPS/W (best-single workload) and \emph{mean} TOPS/W
(workload-equal-weighted) \rev{are listed separately as
supplementary metrics for which higher values are better. For}
chip-level claims we report \rev{the} mean.

\subsection{Experimental Setup}
\label{sec:expt-setup}

\noindent\textbf{Technology node.} All headline results target
ASAP7 7\,nm and use the default DC synthesis calibration described
in Section~\ref{sec:methodology-calibration}, paired with LPDDR5-6400
(40\,pJ/byte). Big and Little tile types run in fixed clock domains
at 1200\,MHz and 500\,MHz, respectively.

\noindent\textbf{Architectures compared.} At each area bracket we
compare three families: \emph{Homogeneous} (\textsf{Homo}, $N$
identical FP16+INT8 tiles), \emph{Big--Little} (\textsf{Hetero-BL},
mixed FP16+INT8 Big and INT4+INT8 Little tiles), and
\emph{Big--Little--Special-Function} (\textsf{Hetero-BLS}, adds Special-Function tiles
for FFT/SNN-integrate/polynomial). The GA refinement
(Section~\ref{sec:expt-dse}) returns \textsf{Hetero-BLS} as best at every
area budget.

\noindent\textbf{Comparison constraints.} Comparisons run under
three constraints: \emph{iso-area} (area matched; report
energy/latency), \emph{iso-latency} (latency matched; report
energy/area), and \emph{iso-energy} (energy matched; report
area/latency).

\subsection{Simulator Validation Setup}
\label{sec:expt-validation}

The simulator is validated along three axes
(Section~\ref{sec:results-validation}): (i) per-module cycle and energy
agreement against Synopsys DC at ASAP7 7\,nm; (ii) end-to-end
agreement against NVIDIA NVDLA at two design points spanning
$32{\times}$ in MAC density (\texttt{nv\_small} 8$\times$8 INT8 64\,KB
CBUF; \texttt{nv\_full} 32$\times$64 INT8+FP16 512\,KB CBUF); and
(iii) a system-level RTL gating study (109 SystemVerilog files,
homogeneous vs.\ iso-area heterogeneous, both synthesized at ASAP7
and run at 1\,GHz with the heterogeneous design's INT4+INT8 tile
power-gated when idle) that grounds the analytical power-gating
model.

\subsection{DSE Setup}
\label{sec:expt-dse}

The DSE pipeline runs two stages, each repeated across multiple
random seeds for confidence-interval estimation.

\begin{table}[!t]
\centering
\caption{\rev{Design-space grid. Asymmetric-precision MAC variants use W$x$A$y$ notation
for $x$-bit weights and $y$-bit activations.}}
\label{tab:dse-knobs}
\setlength{\tabcolsep}{4pt}
\footnotesize
\rev{\begin{tabular}{@{}ll@{}}
\toprule
Knob & Values \\
\midrule
MAC array rows/columns & 8, 16, 32, 64, 128 \\
Per-tile SRAM capacity & 64--4096\,KB in $2{\times}$ steps \\
Precision set & INT8; INT4+INT8; INT8+FP16; \\
 & \quad INT4+INT8+FP16 \\
DRAM bandwidth & 16, 32, 64, 128, 256, 512\,GB/s \\
Tile counts & 1--3 tile types, 1--8 instances per type \\
Sparsity mode & none, activation-sided, two-sided \\
MAC engine type & systolic, spatial, dot-product, CIM \\
Dataflow policy & WS, OS, RS \\
Interconnect & mesh, bus, ring, NoC \\
Double-buffering & off, on \\
Asymmetric-precision MAC & none, W4A8, W2A8, W4A16+W8A16 \\
Pipeline depth & 1, 4, 8, 16 \\
\bottomrule
\end{tabular}}
\end{table}

\noindent\textbf{Design-space grid.} \rev{Table~\ref{tab:dse-knobs}
lists the 12 DSE knobs and the values each can take. 
The winning designs of Section~\ref{sec:results-ga} differ mainly in tile-type composition, supported precision set, Little-tile count, and DRAM bandwidth.}

\noindent\textbf{Sweep.} Stratified random sampling over this
12-knob design space, with three random seeds. Each seed evaluates $\sim$980\,K configurations across
area brackets $\{50, 100, 200, 400, 800\}$\,mm$^2$ and architecture
families \{\textsf{Homo}, \textsf{Hetero-BL}, \textsf{Hetero-BLS}\},
yielding $\sim$2.94\,M total samples. The sweep is the source of
the per-workload best-of-DSE numbers reported in
Section~\ref{sec:results-dse}.

\noindent\textbf{GA refinement.} Per-area-budget GA
(population 200, 100 generations, tournament selection of size 5,
80\% crossover, 20\% mutation, 10\% elitism) seeded from the top 50
individuals returned by the sweep at that budget. Five parallel GA
instances cover the area sweep
$\{50, 100, 200, 400, 800\}$\,mm$^2$. Fitness is the
workload-equal-weighted mean energy savings of the candidate
heterogeneous design over the best homogeneous design at the same
area (found in the sweep), normalized to the best peak TOPS/W
observed. \rev{The DSE engine also supports BO,
compared against the sweep and the GA in Section~\ref{sec:results-ga}.}

\noindent\textbf{Reporting convention.}
Sweep numbers are reported as \emph{cross-seed mean $\pm$ stdev
(95\% confidence interval)} over the three seeds. Per-workload best-of-DSE numbers
report the mean and stdev of the single best sampled design per
workload across the three seeds. GA-refinement results are reported
as the best individual returned by the GA at each
area budget.

\section{Results} \label{sec:results}

The results are organized in three parts.
Section~\ref{sec:results-validation} validates the simulator at three
scales (per-module against ASAP7 synthesis, end-to-end against
NVDLA on a GEMM cross-check, and against synthesized NVDLA-large on
every workload it can execute) and reports the system-level RTL
gating study. Section~\ref{sec:results-dse} presents the multi-seed
DSE per-workload best designs. Section~\ref{sec:results-ga} reports the
GA refinement, the best general-purpose heterogeneous architecture
per area budget, and the three-group workload taxonomy that emerges.

\subsection{Simulator Validation}
\label{sec:results-validation}

\subsubsection{Per-Module Calibration}
The simulator is calibrated against Synopsys DC synthesis at
ASAP7 7\,nm for every RTL module that backs an analytical model,
which is what grounds the absolute energy and area numbers reported
in this paper. SRAM area is taken from CACTI~7.0~\cite{muralimanohar2009cacti}
rather than the standard-cell register array.

\begin{table}[!t]
\centering
\caption{TESSERA vs.\ NVDLA on INT8 $64{\times}64{\times}64$ GEMM,
two design points. \emph{Ratio} columns report TESSERA$/$NVDLA.}
\label{tab:hc-vs-nvdla}
\setlength{\tabcolsep}{3pt}
\renewcommand{\arraystretch}{1.10}
\footnotesize
\begin{tabular}{@{}lrrrrrr@{}}
\toprule
                        & \multicolumn{3}{c}{\textbf{nv\_small} (64 MAC)}
                        & \multicolumn{3}{c}{\textbf{nv\_full} (2048 MAC)}\\
\cmidrule(lr){2-4}\cmidrule(lr){5-7}
\textbf{Metric}         & NVDLA   & TESSERA & Ratio
                        & NVDLA   & TESSERA & Ratio \\
\midrule
Peak TOPS               & 0.064   & 0.064  & $1.00\!\times$
                        & 2.048   & 2.048  & $1.00\!\times$ \\
Latency ($\mu$s)        & 5.12    & 5.52   & $1.08\!\times$
                        & 1.15    & 1.60   & $1.39\!\times$ \\
Energy (nJ)             & 567.7   & 803.1  & $1.41\!\times$
                        & 567.7   & 677.2  & $1.19\!\times$ \\
Area (mm$^2$)           & 0.40    & 0.71   & $1.77\!\times$
                        & 3.31    & 4.96   & $1.50\!\times$ \\
TOPS/W                  & 0.58    & 0.44   & $0.76\!\times$
                        & 4.16    & 4.85   & $1.17\!\times$ \\
\bottomrule
\end{tabular}
\end{table}

\subsubsection{End-to-End vs.\ NVDLA}
We validate end-to-end accuracy against the open NVIDIA NVDLA~\cite{nvdla} at
two design points spanning $32{\times}$ in MAC density:
\texttt{nv\_small} (8$\times$8 INT8 systolic, 64\,KB CBUF) and
\texttt{nv\_full} (32$\times$64 INT8+FP16, 512\,KB CBUF). Both are
exercised on an INT8 $64{\times}64{\times}64$ GEMM that fully fits in
each design's on-chip buffer (Table~\ref{tab:hc-vs-nvdla}).

\noindent NVDLA is the natural---and effectively the only---external
reference for this cross-check: it is the sole openly available
production NPU that ships synthesizable RTL \emph{together} with a
published per-module area and energy breakdown, so it is the one
design an analytical model can be checked against end-to-end at
module granularity. This mirrors established practice for analytical
DSE simulators, each of which is validated against a single such
reference (e.g., Timeloop~\cite{timeloop_ispass2019} against the
Eyeriss accelerator), simply because no second design publishes RTL
and reference numbers at comparable fidelity. The cross-check is
nonetheless stronger than a single comparison: NVDLA is only the
\emph{external} axis of the three-axis validation in this
section---grounded bottom-up by per-module ASAP7 synthesis
(Section~\ref{sec:results-validation}) and topped by a 109-file
system-level RTL gating study---and even that axis spans two design
points across $32\times$ in MAC density, so agreement reflects
scaling behavior rather than a single tuned operating point.

Table~\ref{tab:hc-vs-nvdla} reports each metric and the
TESSERA-over-NVDLA ratio at both scales. \emph{Peak TOPS} matches
by construction. \emph{Latency} ratios ($1.08\times$ small,
$1.39\times$ full) are on par with the fidelity that comparable
analytical simulators (Timeloop~\cite{timeloop_ispass2019},
SCALE-Sim~\cite{scalesim}) report against synthesized references.
\emph{Energy} sits at $1.41\times$ small and shrinks to $1.19\times$
full, a per-inference overhead that amortizes with scale.
\emph{Area} ratios ($1.77\times$, $1.50\times$) reflect a constant
bias: NVDLA's reported area covers only its convolution MAC array and CBUF; once TESSERA
strips its load/store ports and PPM accounting (the
\texttt{nvdla\_cmac}\,$+$\,CBUF subset), the area matches the
synthesized 3.238\,mm$^2$ \textbf{within 2\%} (3.308\,mm$^2$).
\emph{TOPS/W} follows from energy and area. For TESSERA's
\emph{differential} use (comparing architectures under identical
modeling assumptions), a constant bias cancels out, and the
monotone tightening across the $32{\times}$ MAC-density span
between \texttt{nv\_small} and \texttt{nv\_full} confirms scaling
correctness.

\subsubsection{System-Level RTL Gating Study}
At iso-area and 1\,GHz, the synthesized heterogeneous chip
($5{\times}5$ FP16+INT8 tile $+$ $4{\times}4$ INT4+INT8 tile, the
INT4+INT8 tile power-gated when idle) draws \textbf{93.6\% less
power} than the synthesized homogeneous baseline ($2\times$
$4{\times}4$ dual-datapath tiles with the FP16 path clock-gated)
while providing \textbf{28.1\% more MACs} (41 vs.\ 32) in
\textbf{8.3\% less area} (302{,}931 vs.\ 330{,}320\,$\mu$m$^2$). At
500\,MHz the same heterogeneous chip saves 41.9\% area and 70.7\%
power. The 93.6\% power-reduction figure agrees within 6\% of
TESSERA's analytical 95\% leakage-elimination model, providing
direct silicon validation of the gating model. \rev{The study
reports post-synthesis cell area without full place-and-route and
without a Special-Function tile.}

\begin{figure}[!t]
  \centering
  \includegraphics[width=\columnwidth]{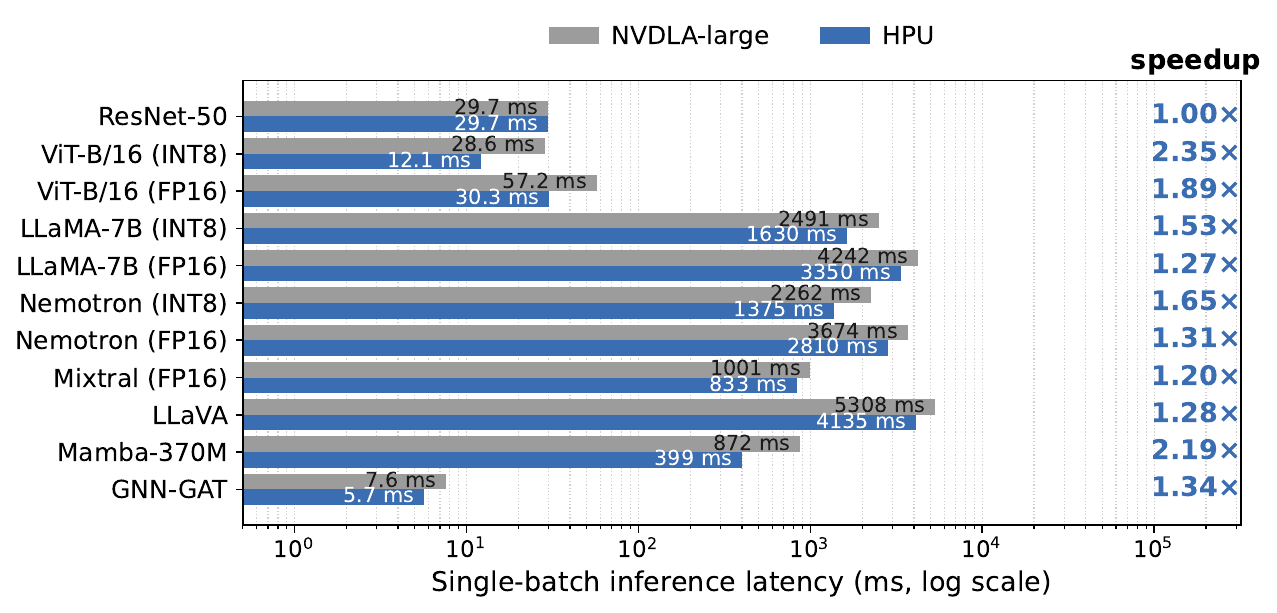}
  \caption{Single-batch inference latency of the GA-refined HPU
  (\textsf{Hetero-BLS}, $\sim$100\,mm$^2$) versus synthesized
  NVDLA-large on every NVDLA-supported workload, with per-workload
  speedup at right.}
  \label{fig:baseline-speedup}
\end{figure}

\subsubsection{End-to-End vs.\ NVDLA-large on the Workload Suite}
\label{sec:results-baseline}
Beyond the GEMM-level NVDLA validation, we compare a representative
GA-refined general-purpose HPU (\textsf{Hetero-BLS},
$\sim$100\,mm$^2$ ASAP7) against the synthesized NVDLA-large
(2048-MAC INT8+FP16 systolic, 512\,KB CBUF) on every workload
NVDLA-large can execute (Fig.~\ref{fig:baseline-speedup}).
\rev{Two findings stand out. First,} the HPU reaches latency
parity on ResNet-50 INT8 (NVDLA's design
target) and is faster on every other NVDLA-supported workload, with
\textbf{$1.5\text{--}2.4\times$ wins on INT8/SSM and compute-bound
ViT workloads} and smaller $1.2\text{--}1.3\times$ wins on the FP16
dense-LLM decodes, where
attention's FP16-only operators serialize on the single Big
tile. \rev{Second,}
the HPU draws $1.1\text{--}2.0\times$ more energy
per inference than NVDLA-large; the energy--latency trade is the
architect's to make and the GA refinement surfaces it as a Pareto
choice (Section~\ref{sec:results-ga}). The four workloads NVDLA-large
cannot execute, namely three INT4 LLMs (LLaMA-7B, Nemotron, Mixtral) and
RT-2's multimodal operators, are omitted from
Fig.~\ref{fig:baseline-speedup}; on them the HPU's INT4-native
Little tile reaches $\sim$0.70\,TOPS/W (RT-2 at 0.27\,TOPS/W),
unlocking deployment of GPTQ/AWQ-style post-training-quantized LLMs~\cite{gptq,awq} that the homogeneous baseline cannot
serve.

\begin{figure}[!t]
  \centering
  \includegraphics[width=\columnwidth]{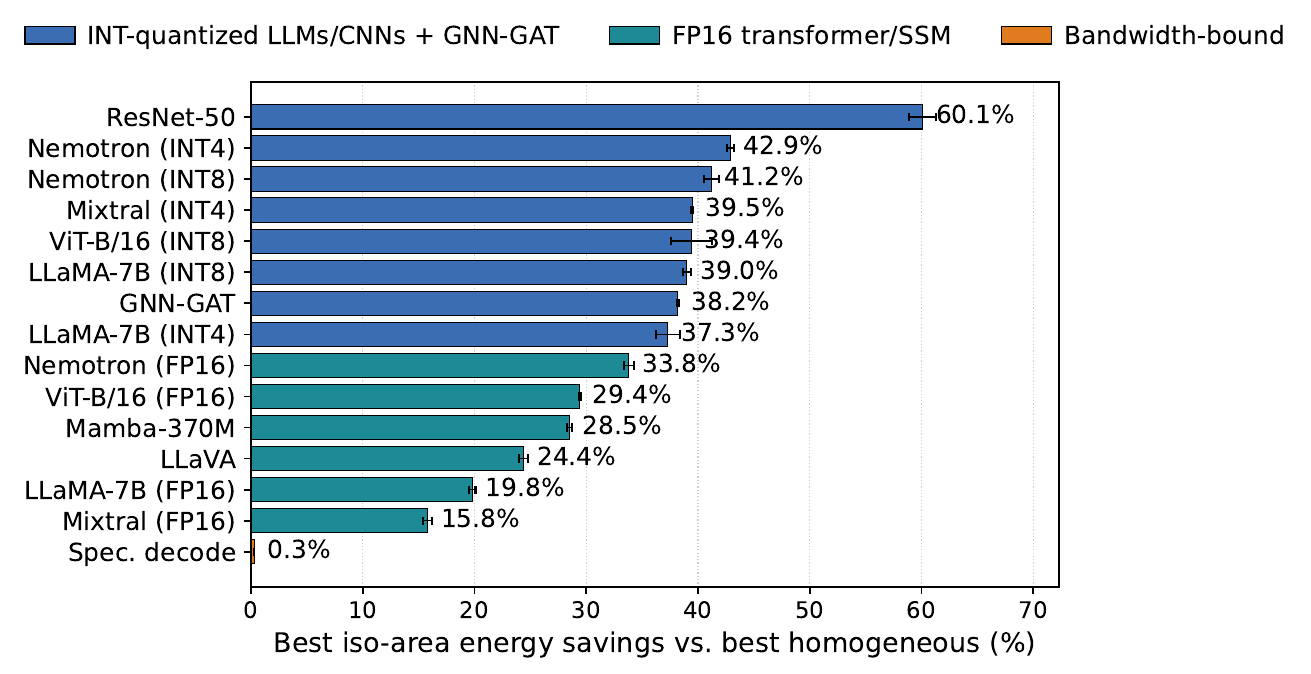}
  \caption{Per-workload best iso-area energy savings of the
  DSE-selected heterogeneous design against the iso-knob homogeneous
  baseline; mean $\pm$ stdev across 3 random-sampling seeds.}
  \label{fig:per-workload-savings}
\end{figure}

\subsection{DSE Across Hardware Knobs}
\label{sec:results-dse}

We ran a multi-seed sweep over the 12-knob design space (3 seeds,
$\sim$2.94\,M samples; Section~\ref{sec:expt-dse}). Each sampled
heterogeneous design is scored by its workload-equal-weighted mean
iso-area energy savings against the iso-knob homogeneous baseline
(Eq.~\ref{eq:dse-objective}). For each workload we record the best
sampled design and report the cross-seed mean/stdev
(Fig.~\ref{fig:per-workload-savings}).

ResNet-50 at $\mathbf{+60.10\pm1.18}$\% is the headline; all
per-workload stdevs sit below 1.82\%. Each bar's winner is
workload-specific: the optimal Big/Little/Special-Function composition is
determined by the target workload's precision and arithmetic
intensity, so the design that wins on ResNet-50 loses on INT4 LLMs
and vice versa. A clear ordering emerges and lines up with the
three-group taxonomy (Section~\ref{sec:results-best}): INT4/INT8 quantized
workloads (plus GNN-GAT) cluster at the top at 37--60\%; FP16
transformer/SSM workloads recover 16--34\%; speculative decoding is
bandwidth-bound (0.28\%). Quantifying the Special-Function tile's
benefit on the five non-MAC workloads (KAN, SNN-VGG9, Hyena, LAVISH,
RT-2) requires an asymmetric per-Special-Function-tile precision
knob that the current symmetric per-tile-precision DSE does not yet
expose; this measurement is reported as future work
(Section~\ref{sec:future-work}). The single-chip general-purpose winner
is reported in Section~\ref{sec:results-ga}.

\begin{figure}[!t]
  \centering
  \includegraphics[width=\columnwidth]{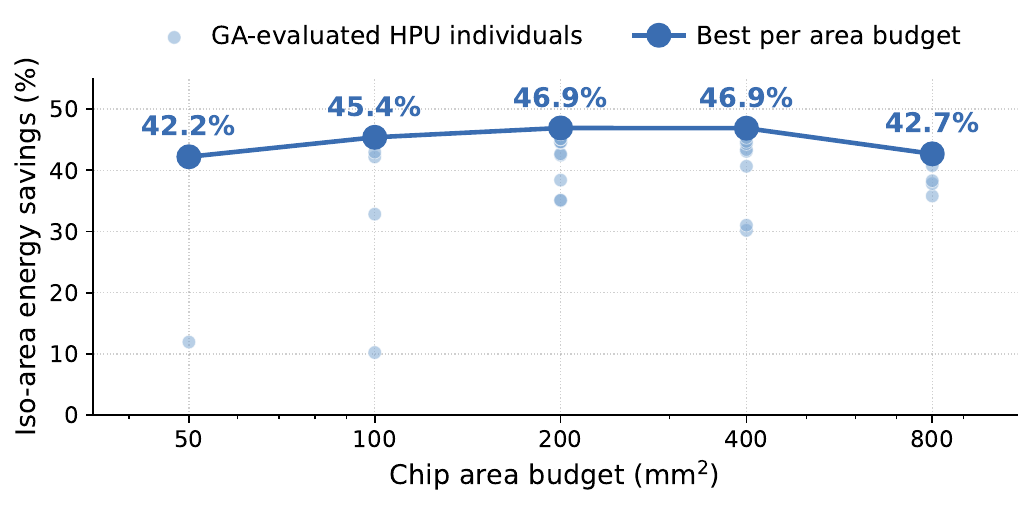}
  \caption{GA-refined mean iso-area energy savings vs.\ chip-area
  budget; the per-budget best traces an inverted-U peaking in the
  100--400\,mm$^2$ band.}
  \label{fig:phase3-pareto}
\end{figure}

\subsection{GA Refinement}
\label{sec:results-ga}\label{sec:results-best}

The GA refinement runs a per-area-budget GA
seeded from the top 50 individuals returned by the random
sweep, scored by the workload-equal-weighted mean iso-area
energy savings of Eq.~\ref{eq:dse-objective}, the same metric
used to score random samples in
Fig.~\ref{fig:per-workload-savings}. The five area instances
evaluate \textbf{28{,}268 GA individuals} in total, with four of
five budgets reaching the ten-generation no-improvement early-stop
(the 50\,mm$^2$ instance terminates against the 144\,h wall after
440 individuals). Fig.~\ref{fig:phase3-pareto} shows the
per-area-budget best alongside the GA-evaluated individuals.
\emph{The GA returns \textsf{Hetero-BLS} as the optimal
architecture at every area budget}: the Big$+$Little$+$Special-Function
composition is robust across an order-of-magnitude span in chip
area.

\noindent\textbf{Best general-purpose chip.} The 200\,mm$^2$
\textsf{Hetero-BLS} winner reaches \textbf{+46.91\% mean iso-area
energy savings} and \textbf{2.44 mean TOPS/W} (peak 10.48), both
workload-equal-weighted across the 20-workload suite. The
inverted-U in Fig.~\ref{fig:phase3-pareto} locates the sweet-spot
in the 100--400\,mm$^2$ band, where the three budgets land within
$1.5$ percentage points of each other ($+45.39 / +46.91 / +46.88\%$);
the optimum is robust to area allocation in this band. The
400\,mm$^2$ design hits the highest mean efficiency at 2.47\,TOPS/W
(peak 11.21). Smaller chips lack room for enough Little tiles, while
larger chips amortize static area over fewer beneficial workloads
(the 800\,mm$^2$ regression to $+42.69\%$ comes from FP16-only ops
that serialize on the few FP16-capable tiles).

\begin{table}[!t]
\centering
\caption{\rev{Search-method comparison at the 200\,mm$^2$ budget.
The GA and BO share the objective, warm
start, and evaluation budget. Runtime is single-worker evaluation time,except the sweep, which ran as 24 parallel jobs. BO seeds span 46.38 to 47.05\%.}}
\label{tab:search-methods}
\setlength{\tabcolsep}{4pt}
\footnotesize
\rev{\begin{tabular}{@{}lrrr@{}}
\toprule
Method & Best savings & Unique configs & Runtime \\
\midrule
Sweep (200\,mm$^2$ bracket) & 18.95\% & 14.1\,K & 2\,h $\times$ 24 jobs \\
Genetic algorithm & 46.91\% & 3{,}253 & 65.0\,h \\
Bayesian opt.\ (median seed) & 46.84\% & 3{,}253 & 75.6\,h \\
\bottomrule
\end{tabular}}
\end{table}

\noindent\textbf{\rev{Search-method comparison.}}
\rev{Table~\ref{tab:search-methods} compares the three search
strategies at the 200\,mm$^2$ budget. The sweep's best
configuration reaches 18.95\% under the objective of
Eq.~\ref{eq:dse-objective}. Seeded identically from the sweep's
top 50, the GA reaches 46.91\% and BO 46.84\%
at the median of three seeds under the same budget of 3{,}253
unique evaluations. BO is the more
sample-efficient of the two, passing 45\% within 500 evaluations
on every seed, about one seventh of the budget. Each method
selects a different \textsf{Hetero-BLS} design of equivalent
quality. The GA winner pairs one $64{\times}8$ Big tile with four
Little tiles and a Special-Function tile, while the BO designs keep one Big and one Special-Function tile but use only one or two Little tiles at higher DRAM bandwidth. 
The plateau near
47\% therefore reflects the design space rather than an artifact
of one optimizer.}

\begin{figure}[!t]
  \centering
  \includegraphics[width=\columnwidth]{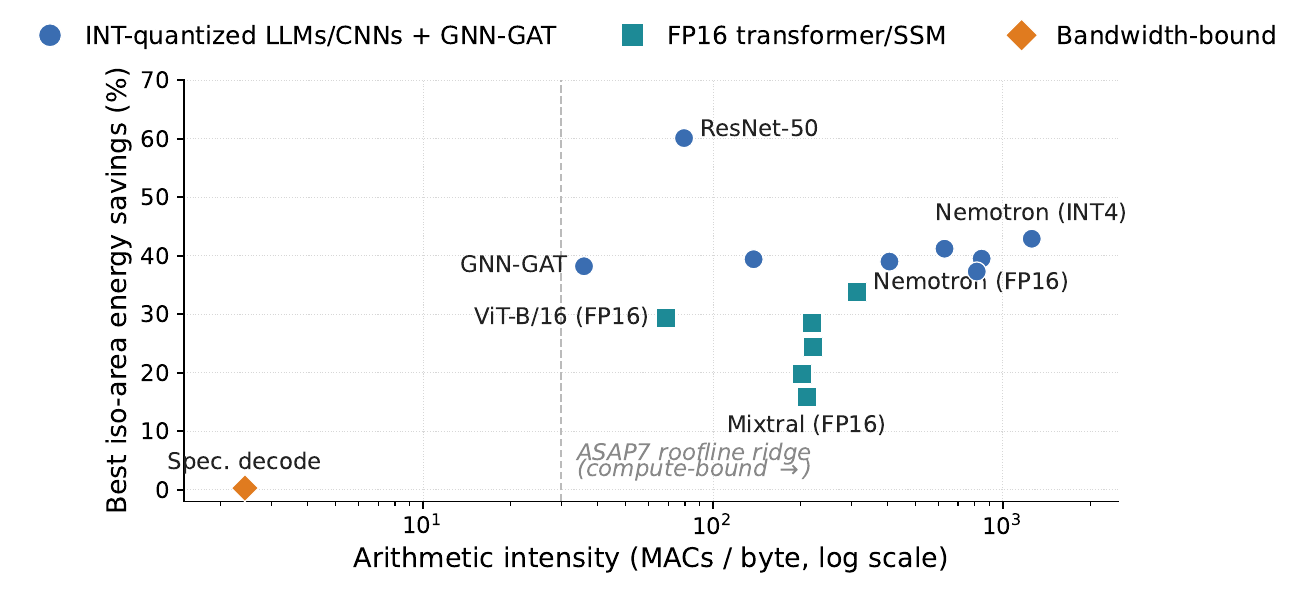}
  \caption{Best iso-area energy savings vs.\ workload arithmetic
  intensity (MACs/byte) for the 15 MAC/DSP-dominant workloads;
  markers colored and shaped by the three-group taxonomy.}
  \label{fig:ai-vs-savings}
\end{figure}

\noindent\textbf{Workload-group taxonomy.}
The 15 MAC/DSP-dominant workloads fall into three groups
(Fig.~\ref{fig:ai-vs-savings}), distinguished by their best
achievable iso-area savings and their position on the
arithmetic-intensity axis. \rev{This grouping is a reading of the
DSE results, not a design input. The design inputs are the tile
types and per-tile knobs, which are motivated by the LNL operator measurements of Section~\ref{sec:bg-diverse-ops}.} \emph{INT-quantized LLMs/CNNs and
GNN-GAT (8)} reach 37--60\% under per-workload tailoring, sitting
at or past the ASAP7 roofline ridge (arithmetic
intensity\,$\geq$\,30\,MACs/byte) where the DSE picks a Big--Little
mix and small INT4+INT8 Little tiles off-load quantizable (and, for
GNN-GAT, INT8-compatible) layers. \emph{FP16 transformer/SSM
workloads (6)} recover 16--34\% in the same compute-bound region
by keeping FP16 attention on the Big tile while off-loading
normalization, gating, and quantizable matmul fragments.
\emph{Bandwidth-bound (1)}, spec.\ decode at an arithmetic
intensity of $2.4$, recovers only 0.28\% because no MAC sizing
helps a memory-starved workload: the single point left of the
roofline ridge in Fig.~\ref{fig:ai-vs-savings}. The remaining five
non-MAC workloads (KAN, SNN-VGG9, Hyena, LAVISH, RT-2) execute on
the Special-Function tile that the GA-returned HPU instantiates;
quantifying their savings against a homogeneous baseline requires
the asymmetric per-Special-Function-tile precision DSE pass
discussed in Section~\ref{sec:future-work}.

\section{Related Work} \label{sec:related_work}

\noindent\textbf{Heterogeneous NPU architectures.}
Prior work has explored heterogeneity along a single axis at a
time. Herald~\cite{herald_hpca2021} and SCAR~\cite{scar2024} compose
sub-accelerators that differ only in \emph{dataflow}
(weight- vs.\ output-stationary), while Stream~\cite{stream_tc2025}
extends this paradigm with layer-fusion scheduling. Tiles in these
designs share identical hardware structures, precisions, and
functional units, leaving the structural efficiency of
right-sized compute unexplored. DNPU~\cite{dnpu_micro2018} pairs a
CNN core with an RNN core in a hard-wired two-core SoC, and the
Renesas AI-MPU~\cite{renesas_isscc2024} integrates a CPU,
reconfigurable processor, and AI accelerator on a 14\,nm SoC; both
demonstrate heterogeneous-core value but are tied to specific
workload pairs and lack any DSE. Spantidi et al.~\cite{spantidi_access}
adopt a big.LITTLE NPU based on MAC precision alone, achieving
$\sim$29\% energy savings, but vary no other knob.
CHARM\,2.0~\cite{charm_trets2024} composes field-programmable
gate-array fabric and AI-Engine
sub-accelerators on Xilinx Versal, and PICACHU~\cite{picachu_asplos2025}
inserts a plug-in CGRA for nonlinear LLM operators next to an
existing GEMM engine; both address complementary aspects of
heterogeneity but do not provide a general-purpose heterogeneous-NPU
substrate or simulator. Maleki et al.~\cite{maleki_hetero} sweep
heterogeneous array sizes alone and report up to 36\% energy
reduction, again on a single axis. \emph{No prior study jointly
varies tile-type composition, precision, sparsity, dataflow, and
special-function-unit presence}, leaving the cross-knob
heterogeneity design space largely unexplored.

\noindent\textbf{Simulation frameworks for NPUs.}
The widely used DSE simulators (Timeloop~\cite{timeloop_ispass2019},
SCALE-Sim~\cite{scalesim}, MAESTRO~\cite{maestro}, and
Voyager~\cite{voyager}) all model homogeneous substrates with a
single tile type, lack dedicated models for non-MAC accelerators
(FFT, SNN-integrate, polynomial), and have no notion of
heterogeneity-aware mapping or DRAM bandwidth sharing across mixed
tiles. As a result, they cannot evaluate the design points that
recent operator-level workload studies on commercial NPUs
(GraNNite~\cite{grannite}, XAMBA~\cite{xamba}, HKN~\cite{hkn})
identify as energy-leaving-on-the-table.

\noindent\textbf{Operator-level orchestration on heterogeneous SoCs.}
A complementary line of work studies how to map individual operators
to the right processing unit on a \emph{fixed} heterogeneous SoC
(e.g., CPU/GPU/NPU pipelines)~\cite{eco,katana}. TESSERA inverts the question:
given a workload mix, it produces the right heterogeneous chip
composition in the first place.

\section[Limitations and Future Work]{\rev{Limitations and}
Future Work} \label{sec:future-work}

\rev{The current DSE does not independently vary precision by tile
type, support within-layer heterogeneity, or use global mapper
look-ahead. The model also omits floorplan and place-and-route
effects, operates at operator-graph rather than kernel
granularity, and models a single NPU rather than a multi-NPU SoC. These limitations define the scope of the reported
46.91\% mean iso-area energy savings, which holds under the
current design space and modeling assumptions. Each item below
states a limitation and the follow-up that addresses it.}

\noindent\textbf{(1) Asymmetric per-tile-type precision DSE pass.}
The GA-returned HPU runs five non-MAC workloads (KAN, SNN-VGG9,
Hyena, LAVISH, RT-2) natively on its Special-Function tile.
Quantifying that tile's iso-area savings needs an asymmetric
per-tile precision knob (e.g., an FP16 SNN integrator alongside
INT4 Big-tile MACs), absent from the current symmetric DSE---our
highest-priority follow-up.

\noindent\textbf{(2) Within-layer heterogeneity.} TESSERA covers
the across-models and across-layers scopes (Section~\ref{sec:bg-three-scopes});
group-level quantization within a matmul or convolution would add
the third, within-layer scope.

\noindent\textbf{(3) Mapper--architecture co-design.} The greedy
DAG mapper lacks global look-ahead, so on sparsely parallel graphs,
tiles with rare op-type compatibility (e.g., the Special-Function
tile during dense-MATMUL stretches) idle. A search-based or learned
policy co-optimized with the tile-mix knobs would recover that slack.

\noindent\textbf{(4) Physical-design feedback.} The area model
omits floorplan dead space, place-and-route (P\&R) complexity, and
power-density hotspots; the RTL gating study (Section~\ref{sec:results-validation})
bounds this mismatch to 8\% on one two-tile composition, with no
physical synthesis at the \rev{7\,nm} target. Floorplan-aware modeling,
full P\&R, and advanced-node synthesis would tighten the iso-area
comparisons and yield manufacturable designs.

\noindent\textbf{(5) Kernel-fidelity simulation.} The simulator
operates at op-graph, not kernel, granularity; fusion savings are
captured via metadata rather than full kernel fidelity. A
kernel-level extension would tighten energy estimates for
fusion-heavy workloads.

\noindent\textbf{(6) SoC-scale heterogeneity across NPUs.} TESSERA
models heterogeneity within a single NPU. Commercial SoCs are
beginning to integrate multiple distinct NPUs on one die---e.g., the
dual-NPU design of MediaTek's Dimensity 9500, which pairs a compute
engine with a CIM core~\cite{dimensity9500_techinsights}. Modeling such
systems requires capturing inter-NPU communication and workload
partitioning across NPUs, promoting TESSERA's chip-level orchestrator
to an SoC-level scheduler.

\section{Conclusion}
\label{sec:conclusion}

\textbf{TESSERA} is the first workload-characterization and DSE
framework that treats structural tile-type composition (Big,
Little, and Special-Function (FFT/SNN/polynomial) tiles, each with a
per-tile configurable precision set) as a joint design space, with an
ASAP7-calibrated simulator cross-validated against NVDLA and a
multi-seed sweep$+$GA pipeline. The GA-refined Big$+$Little$+$Special-Function
HPU reaches \textbf{+46.91\% mean iso-area energy savings} and
\textbf{2.44 mean TOPS/W} on the 20-workload suite (peak
$60.10\pm1.18\%$ on ResNet-50), is \textbf{1.5--2.4$\times$ faster}
than NVDLA-large on INT8/SSM/ViT workloads, and a system-level RTL
gating study confirms \textbf{93.6\% lower power at 28\% more MACs}
in silicon. The contribution is the cross-knob platform that makes
the Big$+$Little$+$Special-Function HPU discoverable in the first
place.


\balance
\def\IEEEbibitemsep{0pt plus 0.5pt}
\bibliographystyle{IEEEtran}
\bibliography{reference}

\end{document}